\documentclass[10pt,aps,prb,amsmath,amssymb,singlecolumn,letterpaper,preprint,
         citeautoscript,showpacs,raggedbottom,balancelastpage,superscriptaddress,longbibliography]{revtex4-1}
\usepackage[usenames,dvipsnames]{color}
\usepackage[pdftex,
  breaklinks,             %
  bookmarks = false, %
  pdfpagemode   = UseNone, %
  pdfstartview 	= FitH,		%
  pdfstartpage  = 1,      %
  colorlinks    = true,   %
]{hyperref}
\hypersetup{
    linkcolor=RubineRed,          
    citecolor=ForestGreen,        
    filecolor=Mulberry,      
    urlcolor=RoyalBlue,           
}
\usepackage{graphicx}
\usepackage[protrusion=true,expansion=true,final]{microtype}
\usepackage{tikz}
\usepackage{longtable}
\usepackage{multirow}
\usepackage{booktabs}
\usepackage{tabularx}
\usepackage{amsmath}
\usepackage{xr}
\DeclareMathOperator*{\argmax}{arg\,max}



\begin{document}

\title{Integrating Machine Learning with Mechanistic Models for Predicting the Yield Strength of High Entropy Alloys}
\author{Shunshun Liu}
\affiliation{\footnotesize{Department of Materials Science and Engineering, University of Virginia, Charlottesville, VA 22904, USA}}
\author{Kyungtae Lee}
\affiliation{\footnotesize{Department of Materials Science and Engineering, University of Virginia, Charlottesville, VA 22904, USA}}
\author{Prasanna V. Balachandran}
\email{pvb5e@virginia.edu}
\affiliation{\footnotesize{Department of Materials Science and Engineering, University of Virginia, Charlottesville, VA 22904, USA}}
\affiliation{\footnotesize{Department of Mechanical and Aerospace Engineering, University of Virginia, Charlottesville, VA 22904, USA}}

\date{\today}

\begin{abstract}
Accelerating the design of materials with targeted properties is one of the key materials informatics tasks. The most common approach takes a data-driven motivation, where the underlying knowledge is incorporated in the form of domain-inspired input features. Machine learning (ML) models are then built to establish the input-output relationships. An alternative approach involves leveraging mechanistic models, where the domain knowledge is incorporated in a predefined functional form. These mechanistic models are meticulously formulated through observations to validate specific hypotheses, and incorporate elements of causality missing from data-driven ML approaches. In this work, we demonstrate a computational approach that integrates mechanistic models with phenomenological and ML models to rapidly predict the temperature-dependent yield strength of high entropy alloys (HEAs) that form in the single-phase face-centered cubic (FCC) structure. Our main contribution is in establishing a quantitative relationship between the HEA compositions and temperature-dependent elastic constants. This allows us to improve the treatment of elastic constant mismatch to the solid solution strengthening effect in the mechanistic model, which is important for reliable prediction of yield strength as a function of temperature in single-phase FCC-based HEAs. We accomplish this by combining Bayesian inference with ensemble ML methods. The outcome is a probability distribution of elastic constants which, when propagated through the mechanistic model, yields a prediction of temperature-dependent yield strength, along with the uncertainties. The predicted yield strength shows good agreement with published experimental data, giving us confidence in applying the developed approach for the rapid search of novel FCC-based HEAs with excellent yield strength at various temperatures.
\end{abstract}

\maketitle

\section{Introduction}
High entropy alloys (HEAs) that form in single-phase face centered cubic (FCC) structure show high ductility and fracture toughness. \cite{Tsai2014, Zhang2014, Gludovatz2014, Jo2017, Li2016, George2019}
However, unlike their body centered cubic (BCC) counterparts, the HEAs that form in the single-phase FCC structure have relatively low yield strength that restricts their application in high temperature environments.
This shortcoming has triggered numerous experimental and computational efforts in the literature to develop strategies based on solid solution strengthening, grain size, and precipitation hardening that will improve the yield strength of FCC-based HEAs\cite{QI2020140056, LUGOVY202195}.
%
%
In \emph{strictly} single-phase alloys, solid solution strengthening and grain size reduction are the two major mechanisms for hindering the dislocation motion and increasing the strength.
Among the two, the solid solution strengthening effect is considered as one of the core properties of HEAs.
Misfit between the atoms in the solid solution originating from the differences (or mismatch) due to atomic sizes, elastic constants, and electronic charge densities are argued as key variables in governing the solid solution strengthening effects in HEAs\cite{WU2014428, LaRosa2019, Oh2019}.
%

%
Fundamental computational approaches based on first principles calculations and molecular dynamics simulations have remained major drivers in the prediction of mechanical properties of HEAs \cite{Tian2014, maiti2016structural, widom2018modeling, feng2021superior, zhang2021prediction, grant2022integrating}.
However, these approaches are computationally expensive and cannot be used to rapidly navigate the vast HEA search space in an efficient manner.
%
%
Inspired by theory and computations, there are several studies in the literature that have attempted to capture the essence of solid solution strengthening mechanism in compact mechanistic models.
These mechanistic models are meticulously formulated through observations to validate specific hypotheses.
Toda-Caraballo and Rivera-D\'{i}az-del-Castillo developed a model for predicting the contribution of solid solution strengthening to the yield strength in HEAs\cite{toda2015}. In their work, the authors extended the Labusch's approach\cite{labusch1970statistical} to multicomponent alloys (including the HEAs) by developing a description of the variable unit cell parameter and atomic size misfit.
Okamoto et al.
uncovered a correlation between average atomic displacements and athermal yield strength (normalized by shear modulus) in three-, four- and five-component alloys\cite{Okamoto2016}. 
Curtin and co-workers have developed mechanistic models that 
include the effects of defect-dislocation interaction and thermally-activated glide to predict the temperature-dependence of yield strength in HEAs\cite{Varvenne2016}. 
This model is based on the effective-medium theory of strengthening for random alloys, which works under the assumption that each alloying element can be regarded as a ``solute'' in an effective ``solvent'' environment that represents the average alloy properties\cite{LaRosa2019}.
%
Oh et al. developed a model, where they argue that the solid solution strengthening is driven by fluctuations in the solute-dislocation interaction energy\cite{Oh2019}. The authors correlate solid solution strengthening to atomic-level pressure, whose origin was linked to the charge transfer between the nearest neighbor atoms in the crystal structure.
In addition, there are also other mechanistic models discussed in the literature that have treated the solid solution strengthening effect in varying degrees of complexities\cite{WALBRUHL2017301, LUGOVY202195}. 
%
%

As an alternative approach to using mechanistic models, many data-driven machine learning (ML) models have also been developed for predicting the yield strength of HEAs\cite{Wen2019, kim2019first, roy2020machine, Xiong2021, bhandari2021yield, wen2021modeling, huang2021machine}.
Unlike the mechanistic models, these ML models are not constrained to specific single-phase FCC and/or BCC HEAs. 
%
Thus, a vast majority of the published ML work combines all kinds of HEAs (single-phase, dual-phase, multi-phase, and mixed-phase) into one data set for establishing the quantitative composition--property relationships.
While some authors have reported data on new experiments to validate the ML model predictions, others have used post hoc model explainability methods to uncover the hidden trends. 

The foundation of this paper is the mechanistic model by Curtin and co-workers that was developed on the basis of the effective-medium theory\cite{Varvenne2016,George2020}. This model takes as input the lattice constants, elastic constants, and misfit volume to predict the temperature-dependence of yield strength of single-phase FCC and BCC HEAs.
All three input quantities exhibit temperature-dependence. However, it is not common in the literature to include their temperature dependence when predicting the yield strength.  
%
%
The objective of this work is to develop a novel computational pipeline that enables rapid prediction of temperature-dependent elastic constants of HEAs, along with the associated uncertainties.
In the literature, several strategies exist to extract and predict the HEA elastic constants\cite{lee2020temperature, vazquez2022efficient, zhang2022composition}.
However, none of them has been shown to predict their temperature-dependence and quantify uncertainties.
One of the simplest and na\"{i}ve approaches involve using the elastic constants of the pure elements and taking linear combination of weighted contribution of the alloying elements in the HEA alloy composition to represent HEA elastic constant\cite{LUGOVY202195}.
While simple, any deviation from linearity will introduce error in the yield strength prediction. Therefore, uncertainty quantification is crucial to reliably interpret the results.
It is also common to use first principles calculations to predict the elastic constants at 0~K\cite{malica2020quasi} and use them to predict the yield strength of single-phase HEAs.
More recently, surrogate ML models have also been built using the first principles calculation data to alleviate the computational cost\cite{lee2020temperature}. 
While the surrogate models can account for uncertainty due to data sampling, they do not capture the temperature-dependence unless this data is explicitly included in the training data set. 
Therefore, use of 0~K elastic constants to predict the high-temperature yield strength may not be optimal especially when the HEAs show strong temperature-dependence of elastic constants.
For instance, experimental data on the shear modulus of CrMnFeCoNi HEA shows a decreasing trend as a function of temperature; it decreases from 85~GPa at 0~K to 68~GPa at 673~K\cite{Varvenne2016,Wu2014,Haglund2015,Laplanche2015}.
%
%
If we use the 0~K elastic constants to predict the high temperature yield strength, then the predictions will be severely over-estimated. %
In addition, if the prediction uncertainties are not quantified then the over-estimated prediction will likely mislead the HEA design. 
More rigorous treatment of temperature-dependent elastic constants exist within the first principles calculations based on the quasi-harmonic approximation\cite{wang2010first} and molecular dynamics simulations\cite{feng2021superior} exists. However, these calculations are expensive.
%

Here, we start from the work of Varshni who proposed a phenomenological model for the temperature-dependence of elastic stiffness constants\cite{varshni}.  
We extract the parameters of the phenomenological model using Bayes' theorem by calibrating the model with the available experimental data on temperature-dependent elastic constants for pure metals and multi-component alloys.
Ensemble support vector regression (eSVR) based ML models are then used to establish the relationship between the chemical compositions and posterior distributions of the Varshni model parameters. Finally, statistical sampling is used to propagate the predicted elastic constant distribution through the mechanistic model, which in turn predicts the temperature-dependent yield strength, along with the uncertainties, for known and unexplored FCC-based HEAs. Comparison of predicted yield strength with published experimental data shows good predictive performance, which gives us confidence in applying the developed approach to rapidly search for novel FCC-based HEAs with excellent yield strength at room and high temperatures. The developed predictive capability can be accessed via an interactive web application (\url{https://adaptivedesign.shinyapps.io/AIRHEAD/}) to guide the design novel HEAs with targeted properties.

\begin{figure}[h!]
    \centering
    \includegraphics[width=150mm]{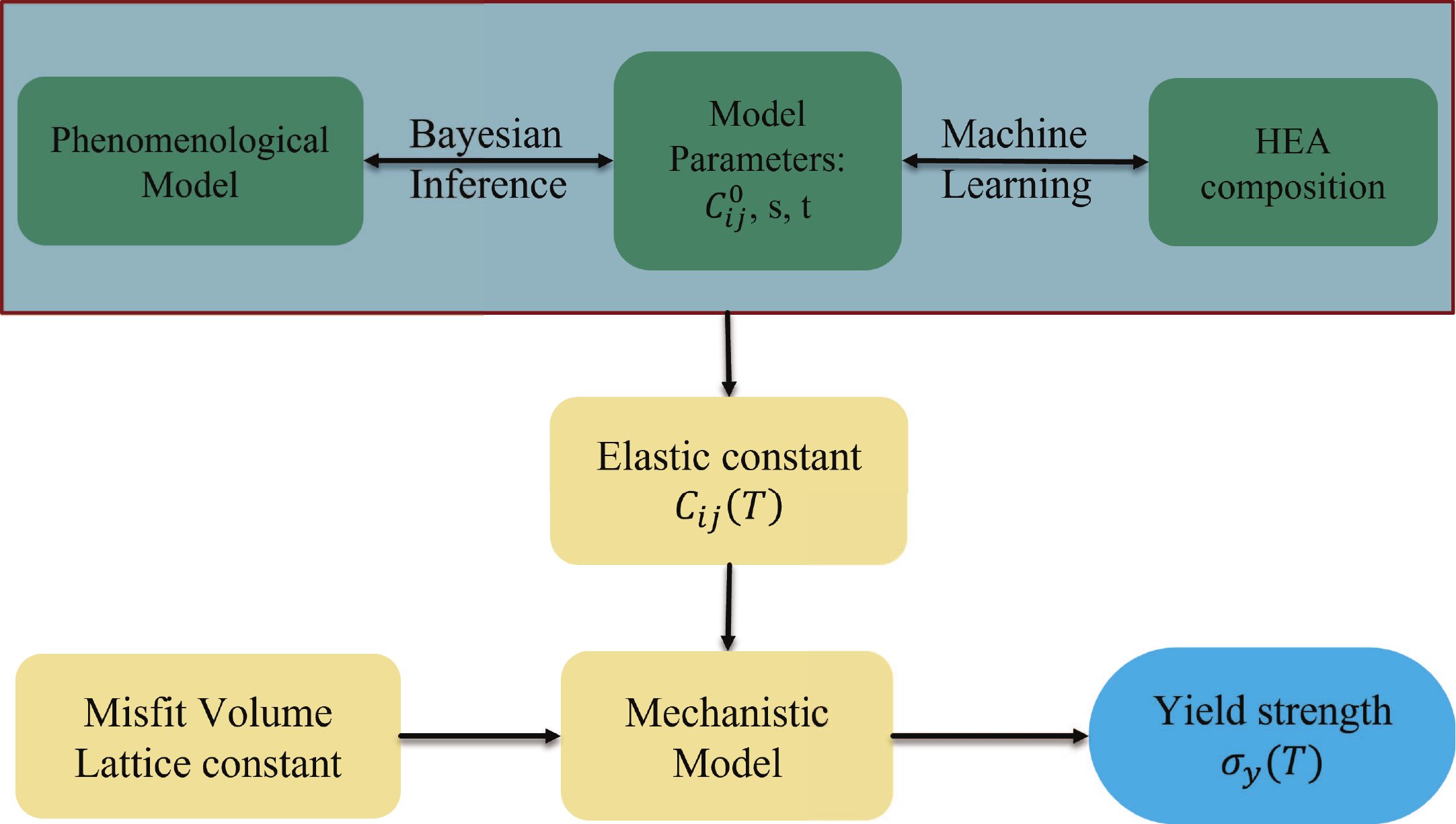}
    \caption{\label{figure:alter_flowchart}The overarching workflow of the computational approach. The green area links the composition to the elastic constants. The output from the Bayesian-Machine Learning (ensemble support vector regression, eSVR) integration is the probability distributions of $C_{ij}$ as a function of temperature [$C_{ij}(T)$]. The misfit volume and lattice constants are considered as deterministic values estimated using Vegard's law from lattice parameter $a$, which was taken from the Materials Project database\cite{Jain2013}. Additional details can be found in the Methods section (\autoref{sec:Methods}).}
\end{figure}

\section{Method}
\label{sec:Methods}
The overarching computational workflow is shown in \autoref{figure:alter_flowchart}. Key components required for the computational work include: (1) Data sets, (2)  Phenomenological Varshni Model, (3) ML, (4) Mechanistic Yield Strength Model, and (5) Integration for HEA design.
%
%

\subsection{Data sets}
We consider three data sets for this work. Data set 1 is an experimental data set of temperature-dependent elastic constants for metals and alloys, whose crystal structures are known. We have surveyed the literature and compiled this data set for 14 materials: 12 of them are pure metals (Al, Cu, Au, Pb, Fe, Mg, Mo, Si, Ag, Ta, Th, and W)\cite{varshni}, one medium entropy alloy (CrCoNi\cite{Laplanche2020}) and one HEA (CrMnFeCoNi\cite{Teramoto2019}). Data set 1 calibrates the Varshni model parameters based on Bayes' theorem. 
%
Data set 2 contains experimentally measured room temperature yield strength data of as-cast FCC-based HEAs. We collected a total of 29 unique FCC-based HEAs, whose as-cast yield strength data are experimentally measured at room temperature\cite{Qin2019,Zhang2019,Sanchez2019,Zhang2019_2,Liu2019,Wu2019,Zuo2014,Wang2009,He2014,Wang2007,Fazakas2015,Laws2015,Tong2005,Ming2017,Jin2019,Qin2019_3,Jiang2019,Niu2019}. 
Not all compositions have equiatomic concentrations of constituent elements.
In addition, experimentally measured room temperature elastic constants data do not exist for the compositions in Data set 2. In Data set 2, we also did not remove the contribution from the grain size effect.
Data set 3 contains experimentally measured temperature-dependent yield strength data for CrCoFeNi and CrCoFeMnNi alloys.
We note that we do not use any of the compositions in Data sets 2 and 3 to calibrate the mechanistic model. Thus, the compiled yield strength data sets (Data sets 2 and 3) are solely used to validate the models and demonstrate the predictive capability of our approach.
All three data sets are given in the Supplemental Material\cite{PROLA_Supp}.

\subsection{Varshni model}
\label{subsec:phenomenology}
In the theory developed by Born and co-workers, it was shown that the temperature-dependence of the elastic constants originate from the variation of the lattice potential energy due to anharmonicity\cite{leibfried1961theory}. 
According to the theory, in the two extreme limits, the lattice contribution to the elastic constants should vary as T$^4$ at very low temperatures and as T at high temperatures.
Varshni, in the year 1970, proposed a phenomenological model for explaining the variation of elastic stiffness constants with temperature\cite{varshni}.
%
The Varshni model can be written as,
\begin{align}
    \label{eq:varshni}
    C_{ij} = C^{0}_{ij} - \frac{s_{ij}}{\exp(t_{ij}/T) - 1}
\end{align}

\noindent where $C^{0}_{ij}$ is the elastic constant at 0~K, $s_{ij}$ and $t_{ij}$ are both fitting parameters, $T$ is the temperature in Kelvin (K) and the subscript $ij$=11, 12 or 44 (since our focus is on cubic materials in this work).
Empirically, for temperatures above $\frac{\Theta_D}{50}$ (where $\Theta_D$ is the Debye temperature), Varshni showed that \autoref{eq:varshni} had the smallest average percentage difference with the experimental measurements. 
For each $C_{ij}$ component, we fit $C_{ij}^0$, $s_{ij}$ and $t_{ij}$ using the experimental data.
While it is common to apply least squares method to extract the parameters of the Varshni model, we use Bayes' theorem for this task (discussed in \autoref{sec:Bayes}) because it also allows us to quantify uncertainties in the extracted model parameters at the expense of a little extra computational cost.
%

\subsection{Bayesian inference}
\label{sec:Bayes}
In Bayesian inference, we use Bayes' theorem\cite{Bayes, Hamada-BayesianBasic, ricciardi2019uncertainty, whalen2021bayesian} to update the parameter distributions in the Varshni's phenomenological model given some experimental data. The mathematical expression of Bayes' theorem can be written as: 
\begin{align}
    \label{eq:bayes}
    P(\theta|\boldsymbol{X}) = \frac{P(\boldsymbol{X} | \theta) P(\theta)}{P(\boldsymbol{X})}
\end{align}
where $P(\theta)$ encodes the prior probability distribution (our perceived knowledge about the model parameters) before the data is collected, $P(\boldsymbol{X} | \theta)$ encodes the likelihood function that describes the joint probability of the observed data as a function of the parameters of the chosen statistical model (Varshni model, \autoref{eq:varshni}), $P(\theta|\boldsymbol{X})$ is the posterior probability distribution that combines the prior distribution and the likelihood function to tell us what information is contained in the observed data, and $P(\boldsymbol{X})$ is a normalizing constant.
%
%
Bayes' theorem (\autoref{eq:bayes}) states that the posterior distribution is proportional to the likelihood function times the prior distribution. To find the value(s) of $\theta$, which is the parameter(s) of the Varshni model, Bayesian inference needs maximum \emph{a posteriori} (MAP) estimation, which can be calculated as follows: 
\begin{align}
    \label{eq:map}
    \hat{\theta}_{MAP} = \argmax_{\theta}P(\boldsymbol{X} | \theta) P(\theta)
\end{align}
Since the likelihood function and prior distributions are generally known, $\theta$ can be estimated using numerical methods.
In this work, we assumed Gaussian priors for the Varshni model parameters ($C_{ij}^0$, $s_{ij}$ and $t_{ij}$) with the following initial values:
$C_{ij}^0 \sim \mathcal{N}(C^0_{lit},50)$, $s_{ij} \sim \mathcal{N}(s_{lit},1)$, $t_{ij} \sim \mathcal{N}(t_{lit},20)$, and a noise $\epsilon \sim \mathcal{N}(0,10)$.
%
The mean values for the priors ($C^0_{lit}$, $s_{lit}$ and $t_{lit}$) are set to be equal to the least square fit from literature (when the data is available). These would constitute an informed prior. 
%
When the least square fit is not available from the literature, the prior distribution was informed by the posteriors of those materials whose mean value are calculated. 
%
The standard deviation of priors for $C_{ij}^0$, $s_{ij}$ and $t_{ij}$ are chosen to be 50, 1, and 20, respectively.
%
Bayesian inference with adaptive Metropolis-Hastings  Markov Chain Monte Carlo (MCMC) algorithm 
is used to find the best match for $C_{ij}^0$, $s_{ij}$ and $t_{ij}$ from the experimental calibration data set\cite{Metropolis1953,mcmc,Haario2001}. 
The outcome from Bayesian inference is a posterior distribution. Our problem formulation results in producing nine posterior distributions.
For each $C_{11}$, $C_{12}$ and $C_{44}$ elastic constant, we have a set of \{$C_{ij}^0$, $s_{ij}$ and $t_{ij}$\}, where ${ij}$ = 11, 12, and 44.
All Bayesian inference calculations are performed using the open source BayesianTools\cite{bayesiantool} package as implemented in the \textsc{R}-language.

\subsection{Machine Learning}
The goal of ML is to establish a quantitative relationship between the HEA composition and the $C_{ij}^0$, $s_{ij}$ and $t_{ij}$ parameters of the phenomenological Varshni model.
We used the dataset (with 14 instances) that was developed for Bayesian calibration to accomplish the ML goal. 
The input descriptors used to represent each metal or an alloy composition in the data set are based on elemental properties\cite{Ward2016}. 
%
%
Initially, we built 36 descriptors and removed redundant descriptors based on the Pearson correlation coefficient (PCC) analysis. Only those descriptors whose pairwise PCC was less than 0.4 was retained. Our PCC analysis down-selected the following three descriptors for ML model building:
\begin{itemize}
    \item ``meanNdValence'' is the weighted average of the number of $d$-electrons in the valence shell taken over all elements in the chemical composition,
    \item ``meanNValence'' is the weighted average of the number of electrons in the valence shell taken over all elements in the chemical composition, and
    \item ``meanAtomicVolume'' is the weighted average of the approximated atomic volume taken over all elements in the chemical composition.
\end{itemize}
%

We used an ensemble of support vector regression (eSVR) models with bootstrap resampling capability to learn the relationship between chemical composition and Varshni's model parameters\cite{svr_tutorial}. 
%
Non-linear radial basis function (RBF) kernel was employed because of its improved generalization ability \cite{Doucet_2007, 1364002}.
Each ensemble was initialized using a modified bootstrap resampling method, where a Monte Carlo sampler was used to take samples from the posterior distribution of Bayesian inference, rather than use the mean value. 
%
We explored a large number of bootstrap resamples (5, 10, 20, 50, 100, 200) to define our ensemble.
%
For each SVR model in the ensemble, a grid search was used to optimize the hyperparameters of the RBF kernel.
We trained a total of nine independent eSVR models to predict $C_{ij}^0$, $s_{ij}$, and $t_{ij}$ for each of the $C_{11}$, $C_{12}$ and $C_{44}$ elastic constants.
%
%
The optimized number of bootstrap resamples for each $C_{ij}^0$, $s_{ij}$, and $t_{ij}$ parameter is listed in Supplementary Table S1.
The final prediction from eSVR is also a probability distribution defined by the ensemble mean and standard deviation.
We utilized the $\epsilon$-SVR method as implemented in the \texttt{e1071} \textsc{R}-package\cite{e1071}.
%

\subsection{Mechanistic Yield Strength Model}
%
The mechanistic model\cite{Maresca2020} for predicting the temperature-dependent yield strength of FCC-based HEAs can be written as, 
\begin{align}
    \label{eq:curtin}
    \tau_{y}(T,\dot{\varepsilon}) &= \tau_{y0}(T)\exp\left[ -\frac{1}{0.55} \left( \frac{k_B T}{\Delta E_b(T)} \ln{\frac{\dot{\varepsilon_0}}{\dot{\varepsilon}}} \right)^{0.91} \right]\\
    \sigma_{y} &= M \tau_{y}
\end{align}

\noindent where the uniaxial yield stress ($\sigma_{y}$) is described as the product of shear stress $\tau_{y0}(T)$ times $M=3.06$, which is the Taylor factor for random FCC alloy. In \autoref{eq:curtin}, $k_{B}$ is Boltzmann constant, $\dot{\varepsilon_0}$ and $\dot{\varepsilon}$ are reference strain rate and experimental strain rate, respectively.  The $\dot{\varepsilon_0}$ is set to $10^4$ $s^{-1}$ and $\dot{\varepsilon}$ is set to $10^{-4}$ $s^{-1}$ according to common practice in literature\cite{Yin2019}. $\Delta E_{b}(T)$ represents the zero-stress activation barrier.
Both $\Delta E_{b}(T)$ and $\tau_{y0}(T)$ are functions of temperature-dependent elastic constants and atomic misfit volume. The $\Delta E_{b}(T)$ is also related to dislocation Burgers' vector. The expression for $\Delta E_{b}(T)$ and $\tau_{y0}(T)$ can be written as\cite{Yin2019}:
\begin{align}
    \label{eq:delEb}
    \Delta E_{b} &= 2.5785\left[ \frac{\Gamma}{b^2} \right]^{\frac{1}{3}} b^3P^{\frac{2}{3}} \delta^{\frac{2}{3}} \\
    \label{eq:tau0}
    \tau_{y0} &= 0.04865\left[ \frac{\Gamma}{b^2} \right]^{-\frac{1}{3}} P^{\frac{4}{3}} \delta^{\frac{4}{3}}
\end{align}

\noindent where $\Gamma$ refers to the dislocation line tension, and the line tension parameter $\alpha$ is set to $1/8$ for FCC alloys; the $\delta$-parameter includes information about effects of the misfit volume:
\begin{align}
    \label{eq:Gamma}
    \Gamma &= \alpha\mu_{111/110} b^2 \\
    \label{eq:delta}
    \delta &= \sqrt{2 \sum_n c_n \Delta V_{n}^2/ (9b^6)}
\end{align}

\noindent For simplicity, we used Vegard's Law (i.e., linear combination of the atomic volume of the constituent elements) to approximate the misfit volume ($\Delta V_{n}$).
We set the Burgers' vector of FCC lattice to be ${a_0}/{\sqrt{2}}$, where $a_0$ is the lattice constant of the HEA. 
We approximated $a_0$ using Vegard's law, i.e., we used linear combination of the lattice constants of the constituent elements to approximate the $a_0$ of the HEA.
Data for atomic volumes and lattice constants are taken from Materials Project database\cite{Jain2013}. These quantities reflect the $Fm\bar{3}m$ space group of FCC metals with some representing the ground state whereas the others in their hypothetical (or metastable) FCC structure. Thus, the misfit volume and Burgers' vector are treated as deterministic parameters, which we acknowledge is a gross approximation that can be improved.
The quantity $P$ in \autoref{eq:delEb} and \autoref{eq:tau0} is a function of elastic constants, which is informed from the combined Bayesian calibration and eSVR modeling pipeline shown in \autoref{figure:alter_flowchart}.
For FCC alloys, $P$ can be written in terms of Voigt-averaged shear modulus ($\mu^{ave}$) and Poisson's ratio ($ \nu^{ave}$):
\begin{align}
    \label{eq:P}
    P(C_{11},C_{12},C_{44}) = \mu^{ave} \frac{1+ \nu^{ave}}{1- \nu^{ave}}
\end{align}

\subsection{Integrated Modeling Workflow}
%
%
%
\autoref{figure:alter_flowchart} shows the integrated modeling workflow that combines the various approaches we have discussed thus far in the Methods section (\autoref{sec:Methods}). Bayesian inference is used to calibrate the phenomenological Varshni model parameters from experimental data and quantify uncertainties in those parameters. The eSVR models are then used to establish the quantitative relationships between HEA compositions and the nine Bayesian inference calibrated model parameters ($C_{ij}^0$, $s_{ij}$ and $t_{ij}$, where $ij$=11, 12 and 44). In total, nine eSVR models are built for predicting the three elastic constants: $C_{11}$, $C_{12}$ and $C_{44}$, along with their associated uncertainties. After establishing the predictive capability for each of the three elastic constants, the probability distribution is propagated to the mechanistic model to give the final probability distribution of temperature-dependent yield strength.


\begin{table}[h!]
\caption{Comparison of least square and Bayesian inference fitting for the CrMnFeCoNi alloy. The column of $C_{12}$ value for the least square fit is left empty because the fit is not available in the literature. The fitting from Bayesian inference contains two values mean and standard deviation (sd), where the least seuqres fitting will only produce the mean values.}
\begin{tabular}{|c|ccc|cccccc|}
\hline
Method    & \multicolumn{3}{c|}{Least Square}                         & \multicolumn{6}{c|}{Bayesian Inference} \\ \hline
Parameter & \multicolumn{1}{c|}{$C_{11}$} & \multicolumn{1}{c|}{$C_{12}$} & $C_{44}$ & \multicolumn{1}{c|}{$C_{11}^{mean}$} & \multicolumn{1}{c|}{$C_{11}^{sd}$} & \multicolumn{1}{c|}{$C_{12}^{mean}$} & \multicolumn{1}{c|}{$C_{12}^{sd}$} & \multicolumn{1}{c|}{$C_{44}^{mean}$} & $C_{44}^{sd}$ \\ \hline
$C^0_{ij}$         & \multicolumn{1}{c|}{202.60} & \multicolumn{1}{c|}{Not Known}    & 138.20 & \multicolumn{1}{c|}{202.64}    & \multicolumn{1}{c|}{1.15}    & \multicolumn{1}{c|}{119.02}    & \multicolumn{1}{c|}{2.14}    & \multicolumn{1}{c|}{137.02}    & 0.62    \\ \hline
$s_{ij}$         & \multicolumn{1}{c|}{25.00}  & \multicolumn{1}{c|}{Not Known}    & 10.10  & \multicolumn{1}{c|}{21.52}     & \multicolumn{1}{c|}{1.38}    & \multicolumn{1}{c|}{8.10}      & \multicolumn{1}{c|}{5.36}    & \multicolumn{1}{c|}{9.84}      & 0.65    \\ \hline
$t_{ij}$         & \multicolumn{1}{c|}{425.80} & \multicolumn{1}{c|}{Not Known}    & 222.10 & \multicolumn{1}{c|}{428.08}    & \multicolumn{1}{c|}{20.35}   & \multicolumn{1}{c|}{872.34}    & \multicolumn{1}{c|}{22.31}   & \multicolumn{1}{c|}{226.11}    & 13.88   \\ \hline
\end{tabular}
\label{tab:ols_bayesian}
\end{table}

\subsection{Global Sensitivity Analysis}
To understand the relative importance of factors that controls the Varshni model and the mechanistic yield strength model, we performed the sensitivity analysis. We utilized the well-known Sobol's method, which is a variance-based method that quantifies how variance in input parameters contribute to the variance of the output\cite{saltelli2002making}.
We calculated the contributions due to the main effect (also known as the 1$^\textrm{st}$-order Sobol index), 2$^\textrm{nd}$-order effect, 3$^\textrm{rd}$-order effect and the total effect\cite{sobol1993sensitivity}. The expression for the 1$^\textrm{st}$-order Sobol index is given in \autoref{eq:1st_order_sobol}, where $V_i$ represents the conditional variance due to factor $i$ and $V(Y)$ represents the unconditional variance of the output. This 1$^\textrm{st}$-order index shows the contribution of a single input factor $i$ to the output variance\cite{saltelli2008global}. Expression for the 2$^\textrm{nd}$-order index is given in \autoref{eq:2nd_order_sobol}, where $V_{ij}$ indicates the variance due to the interaction of factor $i$ with $j$. Thus, the 2$^\textrm{nd}$-order index can capture the variance contribution due to the interaction between two factors.
Similarly, a 3$^\textrm{rd}$-order index is given in \autoref{eq:3rd_order_sobol}, and it indicates the variance due to the interaction between factor $i$, $j$ and $k$.
The total effect index can be calculated using \autoref{eq:total_effect}, where $E(Y|X_{\sim i})$ is the expected value of $Y$ when all factor, except $X_i$, are fixed. The total effect can be interpreted as the total contribution of an input factor from both the 1$^\textrm{st}$-order effect as well as any higher-order interaction terms\cite{homma1996importance}. The Sobol sensitivity analysis is performed using the open source \textsc{sensitivity} package\cite{sensitivity_package} as implemented in \textsc{R}-language.
\begin{align}
    S_i &= \frac{V_i}{V(Y)}  \label{eq:1st_order_sobol}\\ 
    S_{ij} &= \frac{V_{ij}}{V(Y)} \label{eq:2nd_order_sobol} \\
    S_{ijk} &= \frac{V_{ijk}}{V(Y)} \label{eq:3rd_order_sobol} \\
    S_{Ti} &= 1 - \frac{V[E(Y|X_{\sim i})]}{V(Y)} \label{eq:total_effect}
\end{align}

\section{Result and discussion}
%
%
We first use the Bayes' theorem to calibrate the Varshni model parameters ($C_{ij}^0$, $s_{ij}$ and $t_{ij}$) from available experimental data on temperature-dependent elastic constants.
%
%
\autoref{tab:ols_bayesian} shows a representative result of the Bayesian calibration procedure 
for the CrMnFeCoNi Cantor alloy, 
where we compare the outcome with the traditional least squares fit.
Since the least square fits for the $C_{12}$ elastic constant is not given in the literature, we cannot compare it with the Bayesian calibration.
The data shown in \autoref{tab:ols_bayesian} reveals that the mean value from the Bayesian calibration fitting agrees closely with that of the least squares fitting procedure, which is expected.
In addition to the mean value, the Bayesian calibration also provide the uncertainties associated with the mean value. It is important to note that the uncertainties are not readily accessible using the least squares fitting procedure.
%
%
While the uncertainties for the $C_{ij}^0$ and $s_{ij}$ parameters in the Varshni model are fairly small, we consistently find a large uncertainty for the $t_{ij}$ parameter.
%
The source for relatively large uncertainty in $t_{ij}$ can be attributed to any of the following reasons: (1) the nature of Varshni model itself, (2) the prior distribution setup, and (3) the experimental data that we used for Bayesian calibration. In \autoref{eq:varshni}, the parameter $t_{ij}$ appears in the $[\exp(\frac{t_{ij}}{T})-1]^{-1}$ term, whose origin can be traced to Varshni's assumption of Einstein model of solid. Since the posterior mean values for the $t_{ij}$ parameter are close in magnitude (order of $10^2$) to the temperature term in \autoref{eq:varshni}, a standard deviation of $\sim$10-20 on $t_{ij}$ is not expected to have a large impact on the prediction of $C_{ij}$. Since we set the standard deviation in the prior distribution for $t_{ij}$ to 20, the Bayesian calibration procedure did not add or reduce the uncertainty based on MCMC search because there were no experimental elastic constants data where $T \gg t_{ij}$. 

\begin{figure}[h!]
\includegraphics[width=160mm]{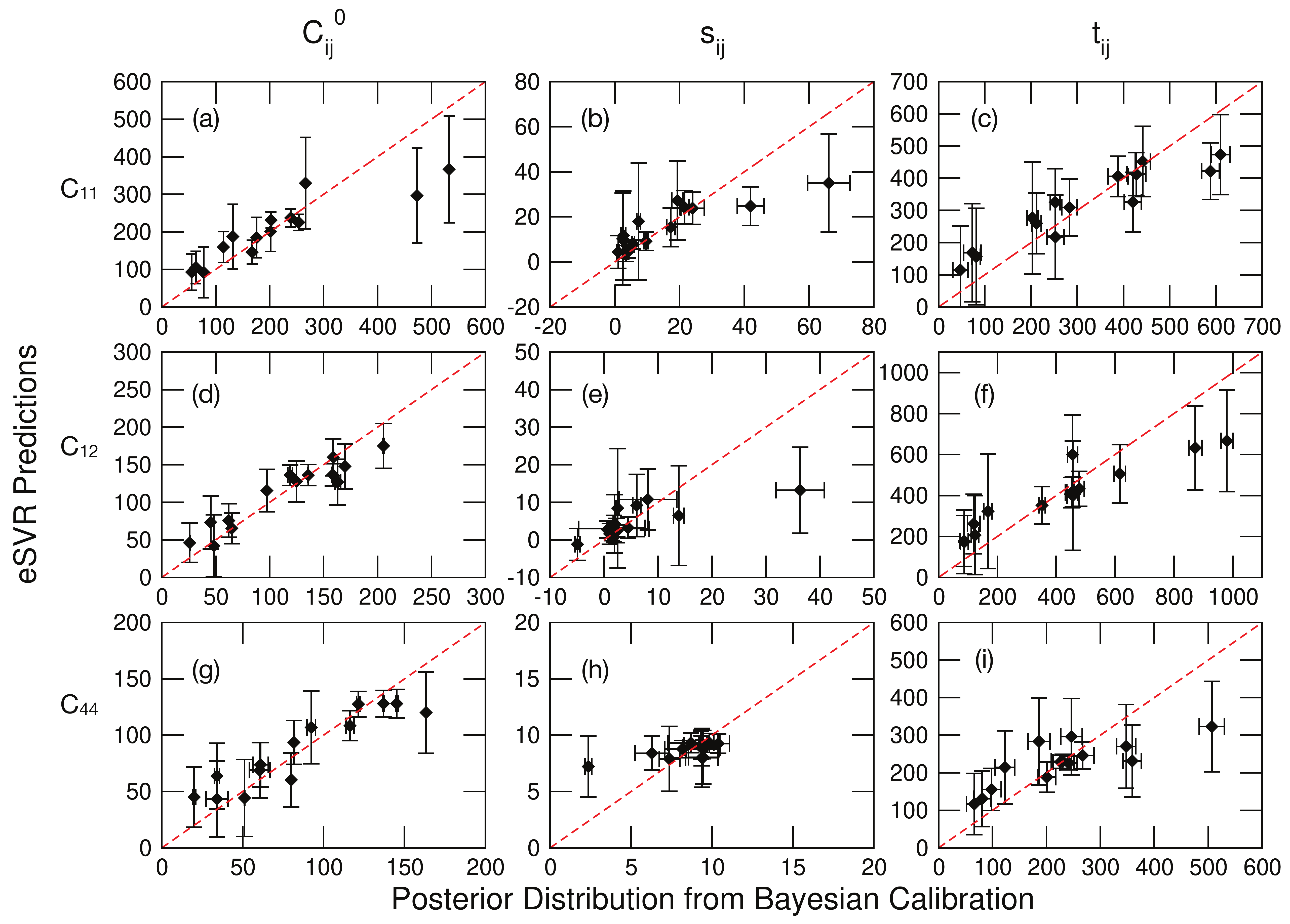}%
\caption{\label{figure:bayesian_ml} Comparison of eSVR predictions (Y-axes) and Bayesian calibrated Varshni model parameters (X-axes). The three rows represent Varshni parameters for $C_{11}$ (top), $C_{12}$ (middle) and $C_{44}$ (bottom). (a) $C_{11}^0$, (b) $s_{11}$, (c) $t_{11}$, (d) $C_{12}^0$, (e) $s_{12}$, (f) $t_{12}$, (g) $C_{44}^0$, (h) $s_{44}$ and (i) $t_{44}$.}
\end{figure}

%
After Bayesian calibration, the next step involves establishing a quantitative relationship between the chemical compositions and the calibrated nine Varshni model parameters from the previous step.
We trained nine independent eSVR models: one for each $C^{0}_{ij}$, $s_{ij}$ and $t_{ij}$ parameter.
For each model, we performed rigorous hyperparameter optimization using a grid search.
The final number of SVR models in the ensemble for each of the parameter is given in \textcolor{blue}{Supplemental Table S1}.
The optimized performances of the nine trained eSVR models are shown in \autoref{figure:bayesian_ml}. The horizontal and vertical axes represent the Bayesian inference calibrated values and eSVR predictions, respectively. These nine plots carry two error bars that reflect the standard deviation of predictions from the Bayesian inference and the eSVR models. 
From \autoref{figure:bayesian_ml}, we infer that the eSVR predictions are relatively more accurate for the $C^0_{ij}$ parameter (have high goodness-of-fit as informed by R$^2$), when compared to that of the $s_{ij}$ and $t_{ij}$ parameters. 
In addition, \autoref{figure:bayesian_ml} shows that the data is not sampled uniformly throughout the search space.
%
Presence of non-uniform sampling indicates the eSVR models did not learn well in the regions of sparse data distribution. Thus, more data is needed in specific regions of the search space for improving the eSVR performance that opens the door for exploring the principles of adaptive materials design\cite{balachandran2020adaptive} to strategically collect data on temperature-dependent elastic constants.

\begin{figure}[h!]
    \centering
    \includegraphics[width=145mm]{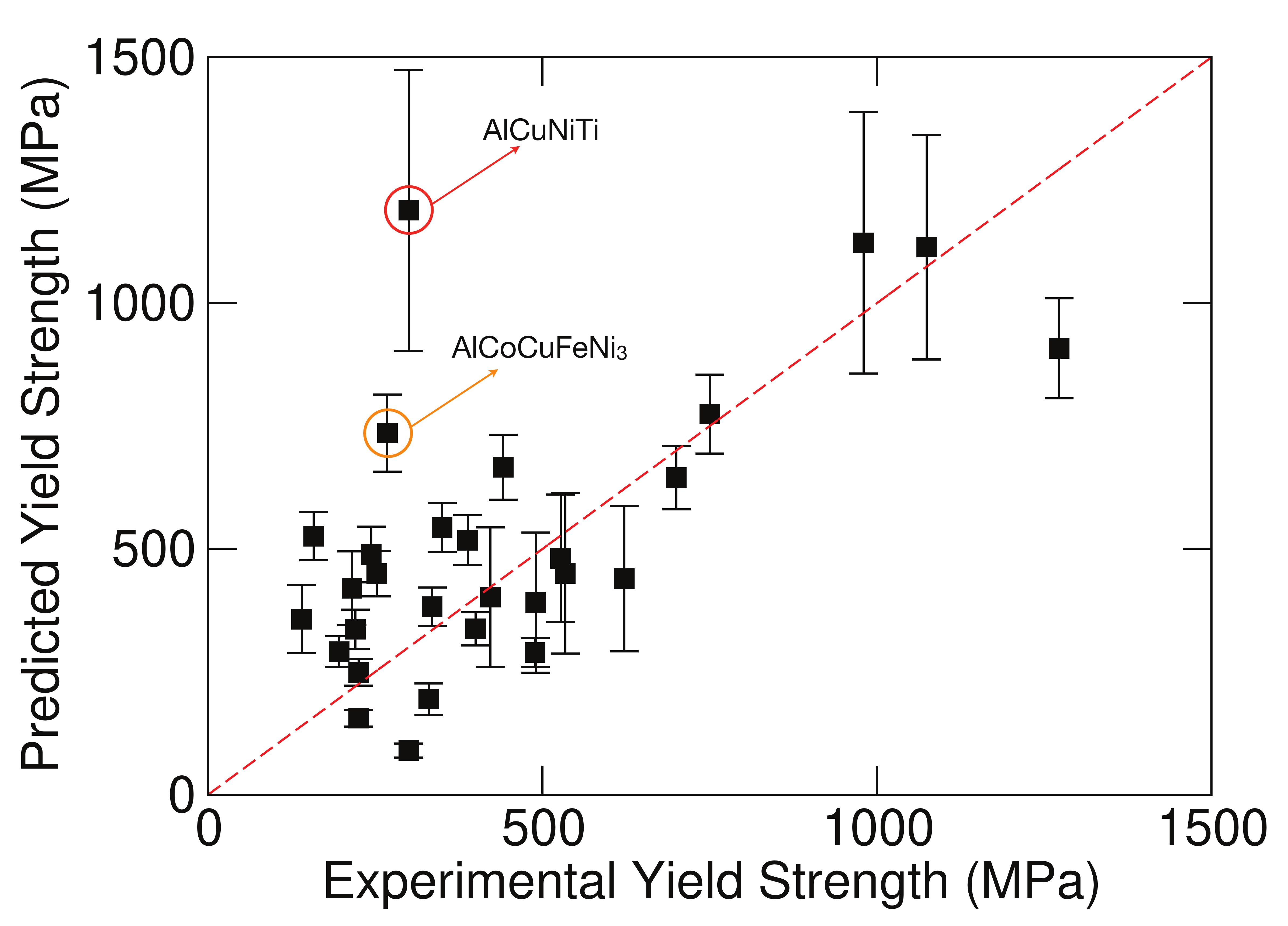}
    \caption{The prediction of yield strength at 300~K for FCC HEAs by the mechanistic model that uses the predicted elastic constants distribution from the Bayesian-eSVR computational pipeline. Most of the data points fall on the X=Y line, indicating good agreement. We circled two outlier HEAs in the plot, where the predictions do not agree well with the experimental data. The upper red circled data point corresponds to that of AlCuNiTi taken from the work of Fazakas et al\cite{Fazakas2015}. The lower orange circled data point corresponds to that of AlCoCuFeNi$_3$ reported by Liu et al\cite{Liu2019}. We then analyze these outliers in detail to explore the plausible reasons for discrepancy.}
    \label{fig:ys_prediction}
\end{figure}

%
The synergy between Bayesian calibration and eSVR models enable rapid prediction of the nine Varshni model parameters, along with the associated uncertainties. 
This then allows us to use the \autoref{eq:varshni} to rapidly predict the temperature-dependent $C_{ij}$ for any arbitrary HEA composition.
%
%
%
%
%
Thus, for each HEA composition, we have a Gaussian distribution of $C_{11}$, $C_{12}$ and $C_{44}$ elastic constants parameterized by mean and standard deviation.
We use this distribution to randomly draw 1,000 samples of $C_{11}$, $C_{12}$ and $C_{44}$.
We then predict the yield strength at 300~K for each of the HEA in Data set 2 using the mechanistic yield strength model (\autoref{eq:curtin}).
Since we have 1,000 $C_{11}$, $C_{12}$ and $C_{44}$ samples for each HEA, we end up with 1,000 predictions of yield strength for each HEA composition.
We then estimate the sample mean and standard deviation from these predictions.
The outcome displaying the yield strength predictions (mean and standard deviation) at 300~K is shown in \autoref{fig:ys_prediction}. Most of the data points fall on the X=Y line, which is encouraging. Thus, for most of the as-cast HEA compositions that form in the single-phase FCC structure, the mechanistic model informed by the temperature-dependent elastic constants can capture the room temperature yield strength trend fairly well. Nonetheless, there are also outliers that appear in \autoref{fig:ys_prediction}, which we discuss next.

The prominent outlier in \autoref{fig:ys_prediction} is the equiatomic AlCuNiTi alloy. Fazakas et al\cite{Fazakas2015} experimentally measured the as-cast yield strength of this alloy as 300~MPa at 298~K, whereas we predict a yield strength as $1188.6\pm 285.87$~MPa from our computational approach. On closer examination of the AlCuNiTi alloy from Fazakas et al.'s paper, we find that the as-cast microstructure is made of the typical dendritic structures. In addition, a strong strengthening effect is observed after annealing, where the experimental compressive yield strength was reported to increase to 1103~MPa (in closer agreement with our predictions). Since Fazakas et al.\cite{Fazakas2015} did not report the room temperature elastic constants for the AlCuNiTi alloy, our predictions cannot be validated. Fazakas et al. reported a value of 7.16\% for the $\delta$ parameter, whereas the Vegard's law underestimates the $\delta$ parameter as 6.7\%. If we use the value of 7.16\% in \autoref{eq:delta}, then the yield strength prediction increases to 1331.02$\pm$311.89~MPa. Furthermore, Fazakas et al. do not provide any insight about the grain size in their paper. Therefore, we were not able to debug the reasons for large discrepancy between our predictions and the experimental result.

The second prominent outlier belongs to the non-equiatomic AlCoCuFeNi$_3$ alloy. Liu et al\cite{Liu2019} experimentally measured the yield strength of as-cast AlCoCuFeNi$_3$ alloy as 268~MPa at room temperature, but we predict a 735.53$\pm$78.55~MPa. In the literature, as-cast AlCoCuFeNi$_3$ was reported to form in single-phase FCC structure. 
One of the sources for this large discrepancy could be attributed to the atomic size difference parameter $\delta$ (appears in \autoref{eq:delta}), which captures the effect of misfit volume. The $\delta$ value reported by Liu et al was 4.92\%, but the Vegard's law provided an $\delta$ estimate of 5.34\%. According to Liu et al, when the relative concentration of Ni is reduced from 3 to 1.5, the $\delta$ value becomes 5.36\% and experimental yield strength increases to 680~MPa at 300~K. However, Liu et al show that the AlCoCuFeNi$_{1.5}$ alloy loses its single-phase identify and forms a mixed-phase made up of FCC+BCC+B2 phases.

\begin{figure}[h!]
    \centering
    \includegraphics[width=160mm]{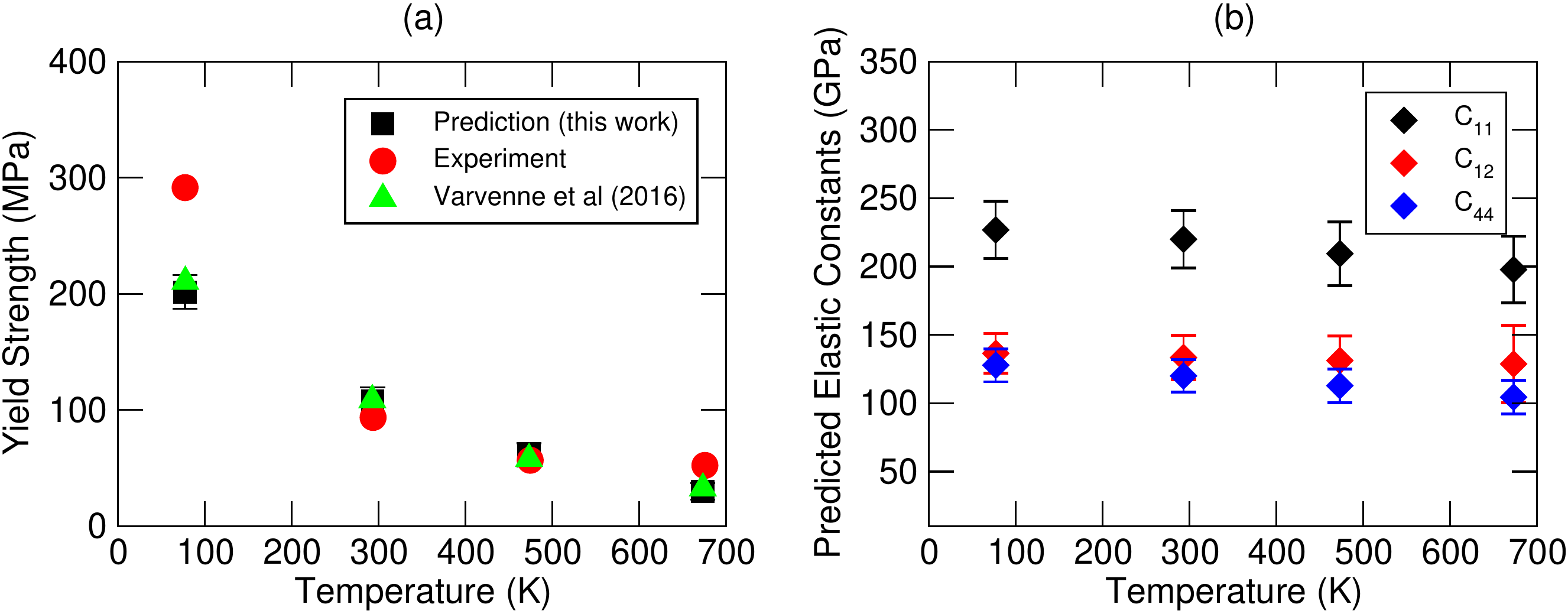}
    \caption{(a) Comparison of yield strength prediction from this work (black), experimental data (red), and theoretical prediction by Varvenne et al. (green) for the four-component CoCrFeNi alloy as a function of temperature. (b) Predicted elastic constants $C_{11}$ (black), $C_{12}$ (red), and $C_{44}$ (blue) for the CoCrFeNi alloy as a function of temperature.}
    \label{fig:ys_prediction_CrCoFeNi}
\end{figure}

We now turn our attention to predictions that involve temperature-dependence of yield strength. 
We focus on three alloys that form in single-phase FCC structure at room temperature: (1) four-component CrCoFeNi alloy, (2) five-component CrCoFeMnNi alloy, and (3) six-component RhIrNiPdPtCu alloy. We predict their yield strengths at 77~K, 293~K, 473~K, and 673~K under the assumption that the crystal structure will remain single-phase in the FCC structure at those temperatures. In the case of CrCoFeNi and CrCoFeMnNi alloys, temperature-dependent yield strength experimental data exists\cite{Varvenne2016}, which gives us an opportunity for validation. However, in the case of RhIrNiPdPtCu, only the room temperature yield strength data is experimentally reported. Our high temperature predictions for RhIrNiPdPtCu remain to be experimentally validated.

In \autoref{fig:ys_prediction_CrCoFeNi}a, we compare the temperature-dependent yield strength prediction with the experimental results and predictions from Varvenne et al\cite{Varvenne2016}. It is known that the mechanistic mechanistic model underestimates the yield strength at 77~K due to the theory neglecting small-scale fluctuations that contribute to the strengthening at zero-temperature and the fact that low temperature prediction is more sensitive to the line tension parameter, $\alpha$\cite{Varvenne2016}. However, the high temperature predictions show good agreement with the experimental data. 
Our yield strength prediction agrees well with the theoretical prediction of Varvenne et al\cite{Varvenne2016} at both low and high temperatures.
%
In \autoref{fig:ys_prediction_CrCoFeNi}b, we show the predicted elastic constants as a function of temperature that was used as an input to predict the yield strength of the CrCoFeNi alloy. All three elastic constants are predicted to show a negative slope with temperature. The $C_{11}$ and $C_{44}$ show a marginally higher temperature-dependence compared to $C_{12}$. 
%
In the work of Varvenne et al, the authors used experimentally measured temperature-dependent shear modulus and Poisson's ratio for making the predictions.
In \textcolor{blue}{Supplemental Table S2}, we compare the experimentally measured temperature-dependent shear modulus for CrCoFeNi with that of the Bayesian-eSVR predictions from this work.
We find excellent agreement between the experimentally measured and predicted values.

\begin{figure}[h!]
    \centering
    \includegraphics[width=155mm]{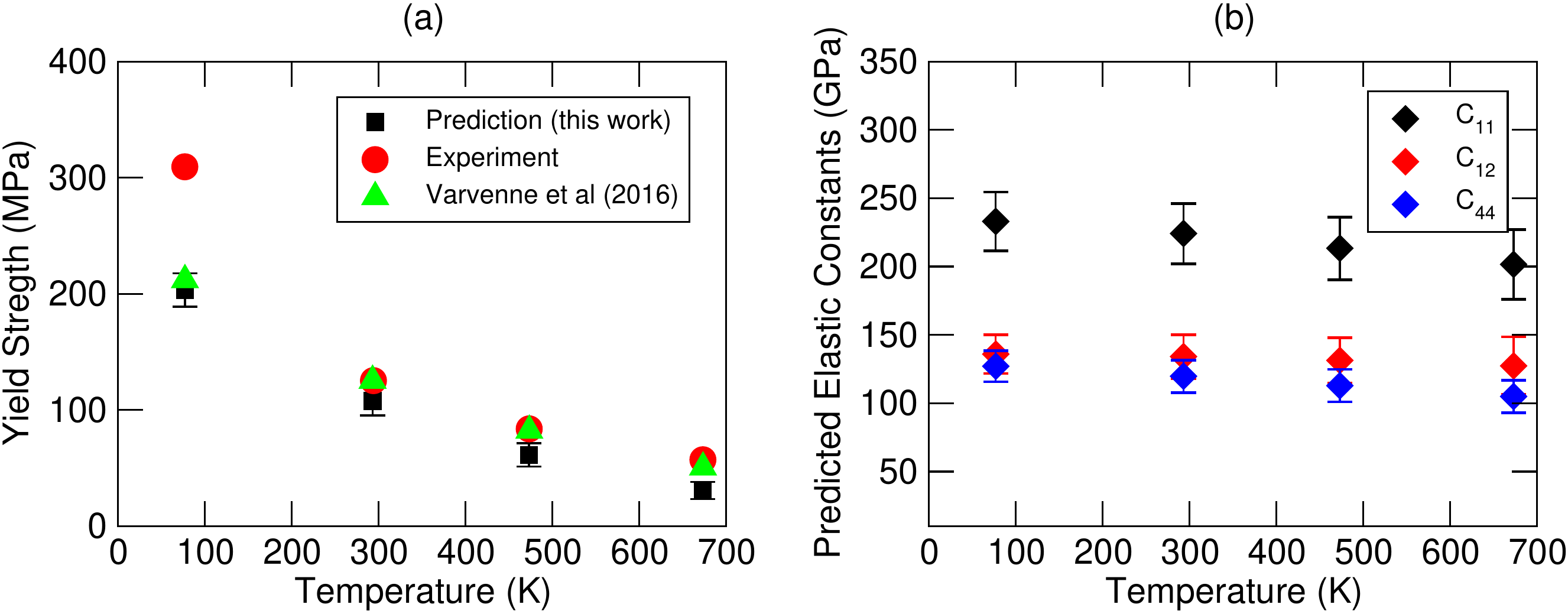}
    \caption{(a) Comparison between yield strength prediction from this work (black) experiments (red), and theoretical prediction by Varvenne et al (green) for the five-component CoCrFeMnNi Cantor alloy as a function of temperature. (b) Predicted elastic constants $C_{11}$ (black), $C_{12}$ (red), and $C_{44}$ (blue) for the CoCrFeMnNi alloy as a function of temperature.}
    \label{fig:ys_prediction_cantor}
\end{figure}

In \autoref{fig:ys_prediction_cantor}(a), we compare the temperature-dependent yield strength prediction for the five-component CrCoFeMnNi alloy with the experimental data and the theoretical prediction of Varvenne et al\cite{Varvenne2016}. Our predicted yield strength is marginally smaller than that of the predictions from Varvenne et al\cite{Varvenne2016}.
Similar to CrCoFeNi, the CrCoFeMnNi alloy also shows a decreasing slope for the elastic constants as a function of temperature (see \autoref{fig:ys_prediction_cantor}b). 
%
In \textcolor{blue}{Supplemental Table S2}, we also compare the experimentally measured temperature-dependent shear modulus for CrCoFeMnNi with that of the Bayesian-eSVR predictions from this work.
Our predictions overestimate the shear modulus at all temperatures, which we attribute as one of the reasons for the marginal discrepancy in the yield strength prediction shown in \autoref{fig:ys_prediction_cantor}(a).
%
%

\begin{figure}[h!]
    \centering
    \includegraphics[width=155mm]{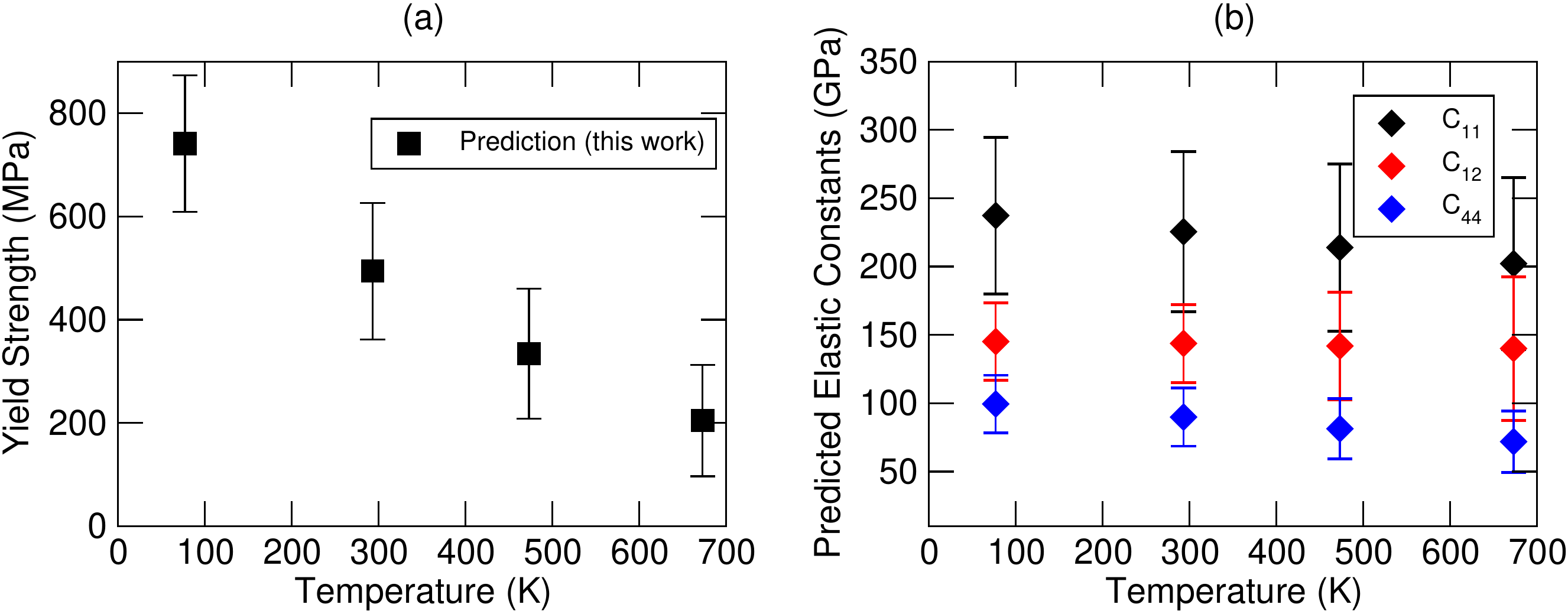}
    \caption{(a) Yield strength prediction from this work as a function of temperature for RhIrNiPdPtCu alloy, whose low and high temperature data do not exist in the literature. (b) Predicted elastic constants $C_{11}$ (black), $C_{12}$ (red), and $C_{44}$ (blue) for the RhIrNiPdPtCu alloy as a function of temperature.}
    \label{fig:ys_prediction_Rh}
\end{figure}

In \autoref{fig:ys_prediction_Rh}(a), we show the predictions for RhIrNiPdPtCu in the temperature values of 77~K, 293~K, 473~K, and 673~K. This six-component HEA was first experimentally reported by Sohn et al\cite{Sohn2017} to form in the single-phase FCC structure. Sohn et al also experimentally measured the compressive yield strength at room temperature and report the value as 527~MPa. Our approach predicted 480.63$\pm$129.54~MPa at 300~K.
The experimentally measured yield strength falls within the error bar of our predictions, which is encouraging.
Yin and Curtin\cite{Yin2019} have also predicted the yield strength using the same mechanistic model (\autoref{eq:curtin}) and report a value of 583~MPa (Yin and Curtin did not provide the prediction uncertainties in their paper).
In addition to validating the 300~K predictions, we also predicted the yield strength at lower and higher temperatures. The results are shown in \autoref{fig:ys_prediction_Rh}(a). We find that the FCC RhIrNiPdPtCu alloy shows higher yield strength than the five-component CrCoFeMnNi alloy system at both low and high temperatures. 
There are two main differences in the application of \autoref{eq:curtin} between this work and that of Yin and Curtin\cite{Yin2019}: (1) Calculation of misfit volume and (2) Calculation of elastic constants.
Yin and Curtin used 0~K DFT calculations to predict the misfit volume and elastic constants.
DFT predicted the $\delta$ parameter as 3.873\%, whereas the Vegard's law predicted $\delta$ as 4.438\%.
In our opinion, the DFT approach is more rigorous than the use of Vegard's law to estimate the misfit volume.
In terms of the elastic constants, Yin and Curtin reported $C_{11}$, $C_{12}$, and $C_{44}$ as 289, 176 and 112 GPa, respectively.
Our predicted temperature-dependent elastic constants is shown in \autoref{fig:ys_prediction_Rh}(b).
Given the inherent limitations within the Varshni model (\autoref{eq:varshni}) in accurately capturing the elastic constants trend at temperatures below $\Theta_D$/50 (see \autoref{subsec:phenomenology} for additional information), we cannot reliably and quantitatively compare our $C_{ij}$ predictions with that of the 0~K DFT work from Yin and Curtin.
Nonetheless, we observe qualitative agreement between our work and that of Yin and Curtin, where the $C_{11} > C_{12} > C_{44}$ trend is preserved. 

Unlike the CrCoFeNi and CrCoFeMnNi alloys, the RhIrNiPdPtCu alloy has relatively large uncertainties in the predicted values of $C_{11}$, $C_{12}$ and $C_{44}$. 
This is because the elastic constants prediction for the RhIrNiPdPtCu alloy is an extrapolation into a new and previously unexplored composition space for our models. 
%
When compared to the CrCoFeNi and CrCoFeMnNi alloys, the RhIrNiPdPtCu alloy has the following similarities and differences in the predicted $C_{11}$, $C_{12}$ and $C_{44}$ elastic constants: (1) The magnitude of the $C_{11}$ value is more or less the same, (2) The mean value of the $C_{12}$ is marginally higher in RhIrNiPdPtCu, (3) The mean value of the $C_{44}$ is marginally lower in RhIrNiPdPtCu.
Thus, we conjecture that the origin of high yield strength in the RhIrNiPdPtCu alloy compared to the CrCoFeMnNi or CrCoFeNi alloys is likely not due to the elastic constant mismatch between the constituent atoms, but due to the atomic size (or volume) mismatch effect.
The predicted low (below 300~K) and high temperature (above 300~K) yield strength data shown in \autoref{fig:ys_prediction_Rh}a and the temperature-dependent elastic constants data shown in \autoref{fig:ys_prediction_Rh}b are reported for the first time in this paper. These results await experimental validation from the community. 
%
%

%
%

\begin{figure}[h!]
    \includegraphics[width=125mm]{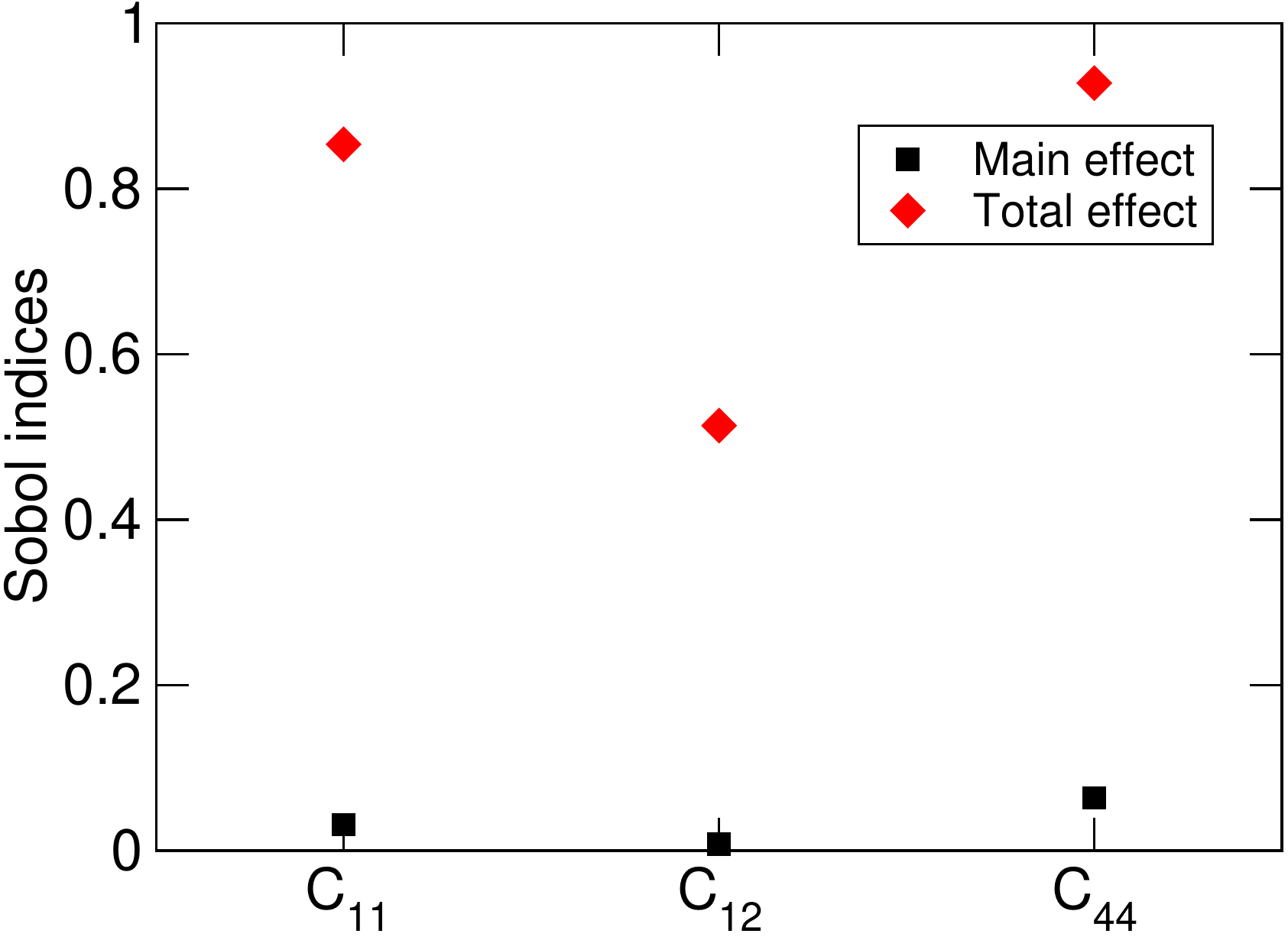}
    \caption{The global sensitivity analysis for the mechanistic yield strength prediction model based on RhIrNiPdPtCu alloy, main effect (black square) is the 1$^\textrm{st}$-order Sobol index ($S_i$) in \autoref{eq:1st_order_sobol} and total effect (red diamond) is the $S_{Ti}$ term in \autoref{eq:total_effect}.}
    \label{fig:sobol_Rh}
\end{figure}

In order to understand the source for large uncertainty in yield strength prediction of the RhIrNiPdPtCu alloy (\autoref{fig:ys_prediction_Rh}a), we performed Sobol's global sensitivity analysis for the yield strength prediction at 300~K. The input parameters are the eSVR predicted $C_{11}$, $C_{12}$ and $C_{44}$ elastic constants at 300~K. 
Although we assumed the $C_{11}$, $C_{12}$ and $C_{44}$ elastic constants to be normally distributed, the Sobol analysis requires samples to be drawn from a uniform distribution $U[0,1]$ hypercube. 
We note that this simplification will not impact the generality of the findings because Sobol showed that any input space can be transformed onto unit hypercube\cite{sobol1993sensitivity}. 
A total number of $2\times 10^6$ samples are drawn to perform the analysis. The Sobol indices are shown in \autoref{fig:sobol_Rh}. The total effect index is much greater than that of the main effect index for all three input parameters, which indicates that the larger variance (or error bar) observed in the yield strength predictions in \autoref{fig:ys_prediction_Rh}a originates from the higher-order Sobol indices (eg., interaction between $C_{11}$, $C_{12}$ and $C_{44}$). 
%
%
The Sobol indices are also calculated for the 300~K yield strength prediction of the five-component CrCoFeMnNi alloy. The result is shown in \textcolor{blue}{Supplementary Figure S1}. We find that the Sobol indices for the CrCoFeMnNi alloy follow the same pattern as that of the RhIrNiPdPtCu alloy (shown in \autoref{fig:sobol_Rh}). The $C_{11}$, $C_{12}$ and $C_{44}$ elastic constants have the same degree of contribution to the variation in the yield strength predictions. 
Thus, the sensitivity analysis reveals that the large uncertainty in the 300~K yield strength prediction seen in the RhIrNiPdPtCu alloy compared to the CrCoFeMnNi alloy is mainly due to the large uncertainty from the eSVR predictions for the $C_{11}$, $C_{12}$ and $C_{44}$ elastic constants.

\begin{figure}[h!]
    \includegraphics[width=125mm]{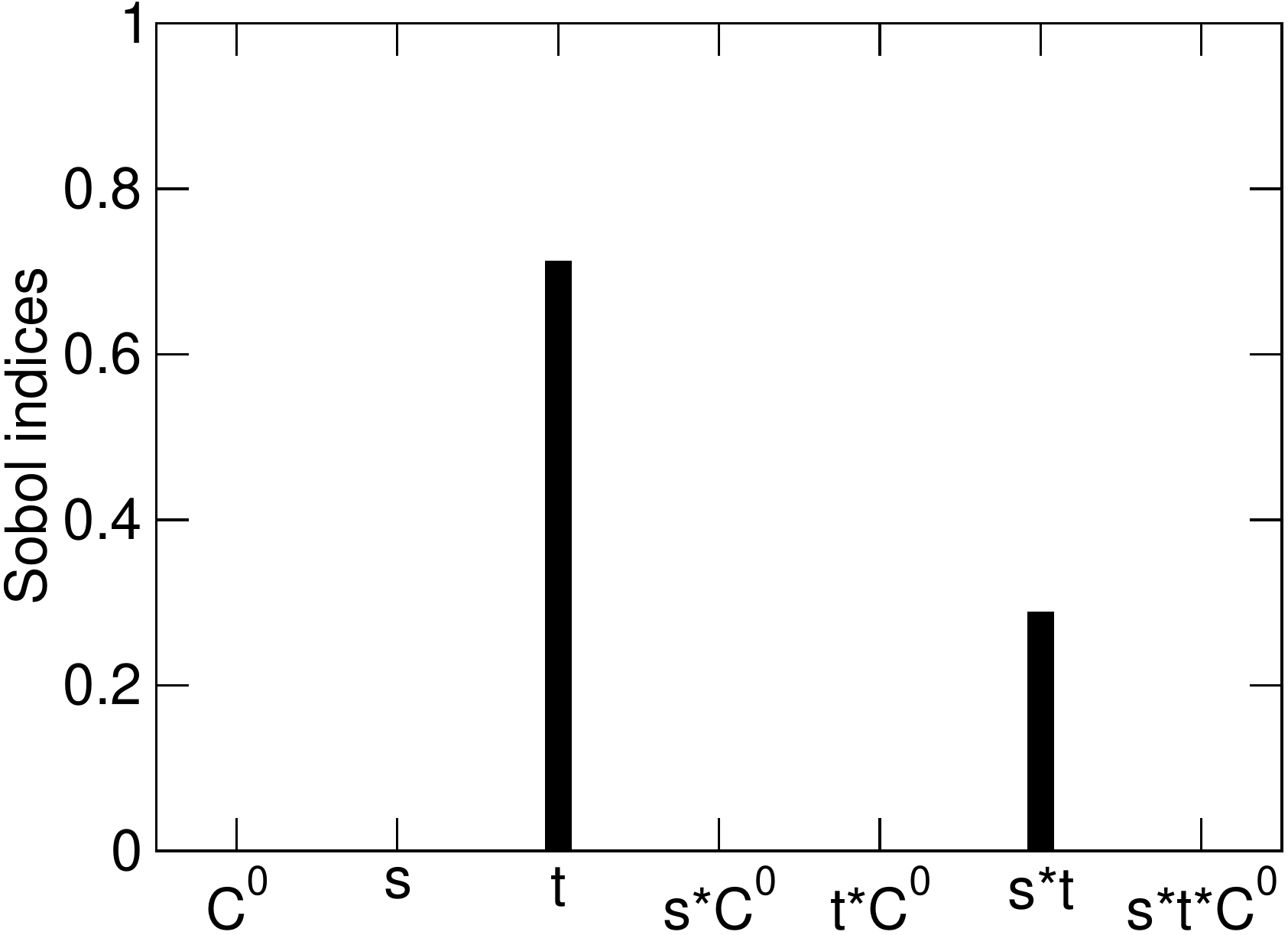}
    \caption{The global sensitivity analysis of Varshni model parameters. The 1$^\textrm{st}$-order Sobol index quantifies the contribution of $C^0$, $s$ and $t$ terms. The 2$^\textrm{nd}$-order Sobol index quantifies the contributions from $s*C^0$, $t*C^0$ and $s*t$. The 3$^\textrm{rd}$-order Sobol index quantifies the contribution from the $s*t*C^0$ term.}
    \label{fig:sobol_varshni}
\end{figure}

We now use the Sobol global sensitivity analysis to understand the source of variance for the predicted $C_{11}$, $C_{12}$ and $C_{44}$ elastic constants from the context of the  Varshni model at 300~K. We use the RhIrNiPdPtCu alloy to discuss the sensitivity analysis. In \autoref{fig:sobol_varshni}, we show the calculated 1$^\textrm{st}$-order, 2$^\textrm{nd}$-order and 3$^\textrm{rd}$-order Sobol index. 
We find that the variance of $t_{ij}$ parameter is the most dominant input factor contributing to the variance of $C_{11}$, $C_{12}$ and $C_{44}$ elastic constants predictions, and the interaction between $s_{ij}$ and $t_{ij}$ is identified as the second dominant contributor. Thus, large uncertainties in the $t_{ij}$ parameters from the eSVR model are expected to contribute to the large variation in the $C_{11}$, $C_{12}$ and $C_{44}$ predictions for the RhIrNiPdPtCu alloy.
For instance, the distributions of $t_{11}$, $t_{12}$ and $t_{44}$ parameters predicted by the eSVR model for RhIrNiPdPtCu is 291.582$\pm$116.198, 330.170$\pm$170.758 and 187.146$\pm$37.51, respectively. The large uncertainties in $t_{ij}$ then propagates to the $C_{11}$ and $C_{12}$ predictions, which invariably results in large error bars. Experimental measurement of temperature-dependent $C_{11}$, $C_{12}$ and $C_{44}$ elastic constants can provide feedback for improving the eSVR predictions.

In an earlier work\cite{lee2022phase}, we built an interactive web application (\url{https://adaptivedesign.shinyapps.io/AIRHEAD/}), where interested readers can input an alloy composition (in the front end) and a trained ML model (in the back end) returns a prediction along with uncertainties about what phase the composition will likely form. We have now updated the web app with new sets of models discussed in the paper, where in addition to phase information the app will also output the predicted elastic constants ($C_{11}$, $C_{12}$ and $C_{44}$) and yield strength at a temperature set by the user. Since the yield strength work discussed in this paper is specific to single-phase FCC HEAs and a particular mechanistic model, we have added a note in the app to inform the users about these restrictions. We believe that web apps provide an excellent avenue to share the research results and the developed predictive capability with the community. This also allows the community to objectively evaluate our approach with the literature data and create a pathway for future improvement through crowd-sourcing.

\section{Summary}
In this paper, we demonstrated a novel computational framework for rapidly predicting the yield strength, along with the associated uncertainties, of single-phase FCC-based HEAs with small experimental data using a mechanistic model. 
The key contribution is the development of a computational approach that shows the integration of a Bayesian calibrated phenomenological model with ensemble ML methods. 
The main outcome is the rapid prediction of temperature-dependent elastic constants of FCC-based HEAs, which is also equipped with the uncertainty quantification capability.
By using these temperature-dependent elastic constants as input to the mechanistic model, we believe that we have improved the contribution from the elastic constant mismatch to the solid solution strengthening effect in single-phase FCC-based HEAs. 
The ultimate impact lies in more reliable prediction of temperature-dependent yield strength in the family of single-phase FCC-based HEAs.
In principle, the developed approach can be extended to predict the yield strength of single-phase BCC HEAs. However, the validity of Vegard's law to capture the $\delta$-parameter (and misfit volume) needs a more critical evaluation.

We now focus on some of the limitations of our approach, which also present future opportunities for improvement.
Our temperature-dependent elastic constants predictions were based on a data set that contained 14 materials.
As more temperature-dependent elastic constants data become available, the predictive capability of our model is expected to improve.
In addition to elastic constants, the mechanistic model also needs the lattice constant and misfit volume as inputs.
We expect both these quantities to depend on the coefficient of thermal expansion, and therefore their temperature-dependence should also be a factor that must be considered for improving the predictive capability.
According to \autoref{eq:tau0} and \autoref{eq:delta}, the final expression of $\sigma_y$ includes the misfit volume raised to the power of $4/3$, which suggests that the misfit volume is the highest order term in the analytical model. This also makes the misfit volume parameter ($\Delta V_n$) one of the most sensitive parameters among others. In this work, the misfit volume is treated as a linear combination of atomic volume from the Materials Project database, which we acknowledge is a gross approximation. 
Another factor that contributes to uncertainty originates from the choice of approximation used to calculate the shear modulus ($\mu^{ave}$) and Poisson's ratio ($\nu^{ave}$).
%
In addition to Voigt average (the method that we use in this paper), there are also alternative methods (eg., Voigt-Reuss-Hill approximation\cite{zuo1992elastic}) that may impact the various terms that appear in \autoref{eq:curtin}.
In this paper, we used the MCMC sampling method for uncertainty propagation because it is fairly straightforward to implement. However, it suffers from low convergence rate. Improving the sampling strategy is also a desired direction for future work.
Thus, there are promising future opportunities that can build upon this work to develop a comprehensive Bayesian framework and predict the properties of HEAs.

\begin{acknowledgments}
Research was sponsored by the Defense Advanced Research Project Agency (DARPA) and The Army Research Office and was accomplished under Grant Number W911NF-20-1-0289. The views and conclusions contained in this document are those of the authors and should not be interpreted as representing the official policies, either expressed or implied, of DARPA, the Army Research Office, or the U.S. Government. The U.S. Government is authorized to reproduce and distribute reprints for Government purposes notwithstanding any copyright notation herein.
\end{acknowledgments}


%

\end{document}